\documentclass[cmp]{svjour}  
\usepackage{latexsym}
\usepackage{amsmath}
\usepackage{amsfonts}
\usepackage{amssymb}
\usepackage{amscd}
\usepackage{amsbsy}
\usepackage{bbm}
\usepackage{fancybox}
\usepackage{pstricks,pst-node}
\usepackage{graphicx}
\usepackage{epsfig}
\usepackage{psfrag}
\usepackage{comment}
\usepackage[normalem]{ulem}

\journalname{Communications in Mathematical Physics}

\newcommand{\sdiffthree}{{\rm SDiff}(S^3)}

\usepackage{float}

\newcommand{\cD}{{\cal D}}

\newcommand{\cL}{{\cal L}}

\newcommand{\cV}{{\cal V}}
\newcommand{\cW}{{\cal W}}

\newcommand{\beq}{\begin{equation}}
\newcommand{\eeq}{\end{equation}}
\newcommand{\bi}{\begin{itemize}}
\newcommand{\ei}{\end{itemize}}
\newcommand{\bt}{\begin{tabular}}
\newcommand{\et}{\end{tabular}}
\newcommand{\bc}{\begin{center}}
\newcommand{\ec}{\end{center}}

\newcommand{\be}{\begin{equation}}
\newcommand{\ee}{\end{equation}}
\newcommand{\bea}{\begin{eqnarray}}
\newcommand{\eea}{\end{eqnarray}}
\newcommand{\ba}{\begin{array}}
\newcommand{\ea}{\end{array}}

\def\bbox{{\,\lower0.9pt\vbox{\hrule \hbox{\vrule height 0.2 cm
\hskip 0.2 cm \vrule height 0.2 cm}\hrule}\,}}
\newcommand{\dsl}{\pa \kern-0.5em /}

\def\tr{{\rm tr}}

\makeatletter \@addtoreset{equation}{section} \makeatother

\def\slashchar#1{\setbox0=\hbox{$#1$}      
  \dimen0=\wd0                 
  \setbox1=\hbox{/} \dimen1=\wd1        
  \ifdim\dimen0>\dimen1            
   \rlap{\hbox to \dimen0{\hfil/\hfil}}   
   #1                    
  \else                    
   \rlap{\hbox to \dimen1{\hfil$#1$\hfil}}  
   /                     
  \fi}

\begin{document}

\title{{Higher spins from Nambu-Chern-Simons theory}}

\titlerunning{Higher spins from Nambu-Chern-Simons theory}

\author{Alex S. Arvanitakis}
\institute{Department of Applied Mathematics and Theoretical Physics, \\Centre for Mathematical Sciences, University of Cambridge, Wilberforce Road, Cambridge CB3 0WA, U.K. \\ \email{A.S.Arvanitakis@damtp.cam.ac.uk}
}
\authorrunning{Alex S. Arvanitakis}
\date{\today}

\maketitle
\hfill DAMTP-2015-66
\vspace{1cm}

\begin{abstract}We propose a new theory of higher spin gravity in three spacetime dimensions. This is defined by what we will call a Nambu-Chern-Simons (NCS) action; this is to a Nambu 3-algebra as an ordinary Chern-Simons (CS) action is to a Lie (2-)algebra. The novelty is that the gauge group of this theory is \emph{simple}; this stands in contrast to previously understood interacting 3D higher spin theories in the frame-like formalism. We also consider the $N=8$ supersymmetric NCS-matter model (BLG theory), where the NCS action originated: Its fully supersymmetric M2 brane configurations are interpreted as Hopf fibrations, the homotopy type of the (infinite) gauge group is calculated and its instantons are classified.
\end{abstract}

\tableofcontents

\section{Introduction}
It is by now well known that interacting 3D higher spin gravity can be formulated without any of the complications of its higher dimensional cousins. This is essentially related to the fact that 3D Einstein gravity (with cosmological constant $\Lambda$) propagates zero degrees of freedom on-shell, and its solutions are always locally flat or $(A)dS_3$. In fact the Einstein action may be written as a Chern-Simons (CS) theory for the isometry group (this was first observed by Ach\'ucarro and Townsend \cite{Achucarro:1987vz}, see also Witten \cite{Witten:1988hc}). When the cosmological constant is nonzero, this is a CS theory for some real form of $\mathfrak{sl}(2) \oplus \mathfrak{sl}(2)$, with the details depending on the sign of $\Lambda$ and whether we are considering Lorentzian or Euclidean signature. 

The first consistent interacting higher spin theories were put forward by Blencowe \cite{Blencowe:1988gj} and Bergshoeff, Blencowe and Stelle \cite{Bergshoeff:1989ns}. These are CS theories defined by a choice of Lie algebra with an $\mathfrak{sl}(2) \oplus \mathfrak{sl}(2)$ subalgebra, and include the familiar $\mathfrak{sl}(n)\oplus \mathfrak{sl}(n)$ theory that describes a finite number of interacting higher spins $\mathbf{2},\mathbf{3} \dots \mathbf{n}$, its infinite dimensional cousin $\mathfrak{sdiff}(S^2)\oplus \mathfrak{sdiff}(S^2)$ (where $\mathfrak{sdiff}(S^2)$ is the Lie algebra of area-preserving vector fields on $S^2$) and various deformations and superextensions thereof. The common feature of this class of theories is that the Lie algebra involved is always a direct sum of the form
\be
\mathfrak{hs}\oplus \mathfrak{hs}
\ee
where $\mathfrak{hs}\supset \mathfrak{sl}(2)$ is a placeholder for any of the aforementioned algebras\footnote{For a more recent higher spin theory of this sort, see \cite{Georgiou:2015tua}.}. Let us note that such a CS theory is parity preserving as long as the action takes the form
\be
S_{CS;\mathfrak{hs}}[A^+]- S_{CS;\mathfrak{hs}}[A^-]\,,
\ee
where both $A^\pm$ are $\mathfrak{hs}$-valued connection 1-forms defined to trade places $(A^+ \leftrightarrow A^-)$ under a parity transformation\footnote{This \emph{intrinsic} parity assignment is on top of the usual transformation of the components of any 1-form under spacetime parity, see e.g. \cite{Arvanitakis:2015oga} where this point is discussed in more detail. This mechanism for parity preservation in Chern-Simons theory was originally introduced in \cite{Hagen:1991ku}.}, and indeed the higher spin theories we have discussed so far are of this form.

In this paper we will propose a consistent theory of higher spin modes interacting with gravity in three dimensions whose gauge group (and algebra) is instead \emph{simple}. That appears to be in tension with parity; indeed the action is not that of an ordinary Chern-Simons gauge theory, but is instead what we will call a \emph{Nambu-Chern-Simons} (NCS) action. This can be seen as a CS theory for the Nambu 3-algebra of smooth functions on the 3-sphere $S^3$ with volume form $\epsilon$, where the ternary Nambu 3-bracket is defined as
\be
\label{nambubracket}
\{ f,g, h\} \equiv \epsilon^{ijk} \partial_i f \partial_j g \partial_k h\,, \quad f,g,h :S^3 \to \mathbb{R}\,.
\ee
While one can, of course, replace the 3-sphere in this definition with another 3-manifold, we will only consider the $S^3$ case in this paper, excepting a few remarks in the discussion later. With this choice of 3-manifold we will refer to this theory as $\sdiffthree$ NCS theory. The gauge group $\sdiffthree$ is the (infinite-dimensional but otherwise ordinary) Lie group of volume-preserving diffeomorphisms of $S^3$. This group is known mathematical literature to be \emph{simple} \cite{Banyaga}\footnote{More precisely, the identity component of $\sdiffthree$ is a simple group; whenever referring to diffeomorphism groups in this paper we will always mean the identity component. The simplicity of $\sdiffthree$ follows from (5.1.3, ii) of the textbook \cite{Banyaga} which asserts that the kernel of the ``flux homomorphism'' $S_\omega$ is simple. For the 3-sphere that kernel is the entire group $\sdiffthree$ because the 2nd de Rham cohomology group vanishes. The first proof of this theorem appears to be \cite{McDuff_sdiff}.}; it appears impossible therefore to write an ordinary CS Lagrangian for this group that preserves parity, and for this reason the NCS Lagrangian cannot be written as an ordinary CS Lagrangian for the corresponding Lie algebra of divergence-free vector fields on $S^3$, $\mathfrak{sdiff}(S^3)$.

A critical role in our analysis will be played by the inclusion
\be
SO(4) \subset \sdiffthree\,.
\ee
A celebrated theorem of Hatcher \cite{Hatcher:1983} asserts that this inclusion \emph{is a homotopy equivalence}; the topology of $\sdiffthree$ is in this sense dominated by that of its $SO(4)$ subgroup\footnote{Hatcher actually proved that $SO(4) \subset {\rm Diff}(S^3)$ is a homotopy equivalence and we thank him for pointing this out; we prove the $\sdiffthree$ version using his original result in Appendix \ref{appendix_instantons}.}. We will find that this rather counterintuitive fact has a field theory counterpart: The $\sdiffthree$ NCS Lagrangian consistently truncates---in a sense we will make precise later---to its $SO(4)$ sector, which is an \emph{ordinary} CS theory. This will allow for an easy proof that the NCS level takes integer values, circumventing the need to extend the arguments of Dijkgraaf and Witten \cite{Dijkgraaf:1989pz} to the non-compact, infinite dimensional case.

The group $SO(4)\cong SU(2)\times SU(2)/\mathbb{Z}_2$ is the gauge group corresponding to the spin 2 sector. Truncating to this sector gives Euclidean gravity on 3D de Sitter space \cite{Banados:1998tb}; likewise, $\mathfrak{su}(n)\oplus \mathfrak{su}(n)$ CS theory describes (Euclidean de Sitter) higher spin gauge theory (see appendix C of \cite{Basu:2015exa}). The $\sdiffthree$ NCS theory therefore also describes higher spins on Euclidean de Sitter space, which we verify by calculating the spectrum of perturbations around this vacuum using harmonic analysis. The gauge field modes naturally organise themselves into an infinite pyramid, where each line is irreducible under $SO(4)$:
\begin{align}
\label{pyramid_of_spins}
\text{spin content} =& \mathbf{2} \oplus \phantom{3} \nonumber \\
                   \:& \mathbf{2} \oplus \mathbf{3} \oplus \phantom{3} \nonumber \\
                   \:&\vdots \nonumber \\
                   \:& \mathbf{2} \oplus \mathbf{3} \oplus \dots \oplus \mathbf{n} \oplus \phantom{3} \nonumber \\
                   \:& \vdots \phantom{\oplus \mathbf{3} \oplus \dots \oplus \mathbf{n} \oplus \phantom{3}\oplus}\ddots
\end{align}
In this sense, the $\sdiffthree$ NCS theory unifies all $\mathfrak{su}(n)\oplus \mathfrak{su}(n)$ CS higher spin theories at once. Our result therefore extends the older results of Bergshoeff, Blencowe and Stelle \cite{Bergshoeff:1989ns} on the smaller, \emph{area}-preserving groups ${\rm SDiff}(M^2)\times {\rm SDiff}(M^2)$.

The NCS action we will consider in this work was originally written down by Bagger and Lambert \cite{Bagger:2007vi} and studied in the more general context of $\rm SDiff$ gauge theory by Bandos and Townsend \cite{Bandos:2008jv} as part of the $\sdiffthree$ NCS-matter model with maximal $N=8$ supersymmetry proposed to describe an---infinite, in this case---collection of parallel M2 branes in M-theory. This lies in the more general class of $OSp(8|4)$-symmetric\footnote{This superconformal symmetry refers to the symmetry of the (N)CS-matter model where the spacetime is nondynamical Minkowski 3-space. In this instance the (N)CS gauge field does not encode the spacetime geometry. This dual interpretation of the NCS Lagrangian is in line with the interpretation of $\mathfrak{sl}(2)\oplus \mathfrak{sl}(2)$ CS as either gravity or gauge theory.} models of Bagger-Lambert \cite{Bagger:2006sk,Bagger:2007vi,Bagger:2007jr} and Gustavsson \cite{Gustavsson:2007vu} (BLG). A BLG Lagrangian is defined by a choice of 3-
algebra and a single dimensionless coupling constant, which appears in the action as a pre-factor. Assuming the 3-algebra is finite dimensional, the requirement of manifest unitarity uniquely fixes the model uniquely \cite{Papadopoulos:2008sk,Gauntlett:2008uf}; the Lagrangian is that of an ordinary $\mathfrak{su}(2)\oplus \mathfrak{su}(2)$ CS-matter theory. The coupling constant is identified with the CS level $k$ and is quantized in the standard way; this integer, along with the choice of gauge group (e.g. $SO(4)$ versus $SU(2) \times SU(2)$), characterises the geometry probed by the M2 branes.

This conclusion was arrived at in the original references \cite{Lambert:2008et,Distler:2008mk,Lambert:2010ji} (see \cite{Bagger:2012jb} for a review) through an analysis of the space of vacuum configurations preserving all supersymmetries. For the $\mathfrak{su}(2)\oplus\mathfrak{su}(2)$ models, the most straightforward approach leads to a nonsensical result that both contradicts expectations from supersymmetry and fails to have an interpretation in terms of M2 branes on some spacetime \cite{VanRaamsdonk:2008ft}. The correct calculation appears to involve careful consideration of global features of the gauge group as well as the presence of instantons \cite{Lambert:2008et,Distler:2008mk,Lambert:2010ji,Bagger:2012jb}. While we were able to prove that the NCS level $k$ of the $\sdiffthree$ theory is quantized, calculating the vacuum moduli space in the manner of \cite{Lambert:2008et,Distler:2008mk,Lambert:2010ji,Bagger:2012jb}---and thereby relating $k$ to the specifics of the geometry probed by the M2 brane 
condensate---faces a technical obstruction. However, the ``na\"ive'' approach goes through for the $\sdiffthree$ BLG model, giving
\be
{\cal M}=\frac{{\rm Embeddings}(S^2 \to \mathbb{R}^8)}{{\rm Diff}(S^2)} \times \mathbb{Z}^*
\ee
which is simply a 2-sphere's worth of indistinguishable planar M2 branes on flat $\mathbb{R}^{2,1}\times \mathbb{R}^8$ times a nonzero integer; we will see that both factors can be understood together geometrically as embeddings of the base of the Hopf fibration $S^3 \to S^2$ in $\mathbb{R}^8$, where the integer is the Hopf invariant.

Finally, we give calculations in appendix \ref{appendix_instantons} of certain topological data of $\sdiffthree$ BLG theory, including the homotopy type of $\sdiffthree$ as well as that of the unbroken gauge group $H$, and we use the results to obtain the topological classification of supersymmetric $\sdiffthree$ instantons. We expect this information will be relevant in future studies of $\sdiffthree$ BLG gauge theory in general, and its vacuum moduli space in particular.

\section{The ${\rm SDiff}(S^3)$ Nambu-Chern-Simons Lagrangian. Harmonic expansion}
The fundamental object of investigation in this paper is going to be the NCS action associated to the infinite 3-algebra of functions on a 3-sphere $S^3$ equipped with the Nambu bracket. This is a gauge theory for the group $\rm SDiff(S^3)$ of volume-preserving diffeomorphisms of $S^3$. A detailed exposition of $\rm SDiff$ gauge theory may be found in \cite{Bandos:2008jv}. Let us mention here that we will not actually use 3-algebra notation, and will instead follow \cite{Bandos:2008jv} in writing down expressions adapted to ${\rm SDiff}(M^n)$. Another paper which treats ${\rm SDiff}$ gauge theory is the review \cite{deAzcarraga:2010mr}, where 3-algebra expressions can be found.

To define ${\rm SDiff}(S^3)$, we introduce a volume form $\epsilon_{i j k}$ on $S^3$ or (equivalently) a nowhere vanishing scalar density $e$. An infinitesimal $\sdiffthree$ gauge transformation is then given by the action of an $S^3$ vector field (and spacetime scalar field) $\xi^i$ that preserves the volume form, or (equivalently) is {\it divergence-free}:
\be
\label{divfree}
  ({\cal L}_\xi \epsilon)_{i j k}=0 \iff \partial_i\left( e \xi^i \right)=0\,.
\ee
The vanishing divergence condition (\ref{divfree}) can be solved for in favour of an unconstrained gauge parameter $\omega_i$ defined by ($\varepsilon^{i j k}$ is the {\it invariant} tensor density)
\be
e \xi^i \equiv \varepsilon^{i j k} \partial_j \omega_k
\ee
up to an $S^3$-exact 1-form which will drop out of all expressions. As a side note, the requirement that we can write any divergence-free vector field in this way forces us to consider 3-manifolds of trivial 2nd de Rham cohomology, singling out the 3-sphere as the simplest candidate $M^3$. The $\sdiffthree$ gauge potential (where $\sigma$ labels the $S^3$ coordinates and $x$ the 3D spacetime ones)
\be
s_\mu^i=s_\mu^i(x;\sigma)
\ee
is an $S^3$-vector-field-valued spacetime 1-form, and is also divergence-free in its $S^3$ index; we will solve the corresponding constraint in terms of an $S^3$ and spacetime 1-form $A_{\mu i}$:
\be
e s^i \equiv \varepsilon^{i j k} \partial_j A_k\,.
\ee

We can now write down the $\sdiffthree$ NCS Lagrangian. If we parameterise $S^3$ by coordinates $\sigma^i, \, i =1, 2,3$ the Lagrangian can be given as an integral:
\be
\label{sdiffcsaction}
L_{\sdiffthree}=\oint d^3 \sigma e \left[ ds^i \wedge A_i - \frac{1}{3} s^i \wedge s^j \wedge s^k \epsilon_{i j k} \right]\,.
\ee
This is invariant up to a spacetime exact term under the gauge transformation
\begin{align}
\delta_\xi s^i&=  d\xi^i - [\xi,s]^i,\    &\delta_\xi A_i=& d \omega_i -\cL_\xi A_i\,,\\
&= d\xi^i -(\xi^j \partial_j s^i-s^j\partial_j \xi^i),  &=& d\omega_i - \xi^jA_j -\partial_i\xi^j A_j\,.
\end{align}
In these expressions we use form notation for spacetime and index notation for $S^3$; $L_{\sdiffthree}$ is an $S^3$-independent spacetime 3-form. We emphasise that despite the ambiguity in the definition of $A_i$, $L_{\sdiffthree}$ is well-defined. The action is of course given as the spacetime integral of the Lagrangian 3-form:
\be
\label{sdiffaction}
S_{\sdiffthree}= \int L_{\sdiffthree}
\ee
This is \emph{parity invariant}: a \emph{spacetime} parity transformation $(x^0,x^1,x^2) \to (x^0, -x^1,x^2)$ can be compensated by a corresponding \emph{internal} $S^3$ parity transformation, e.g. $(\sigma^1,\sigma^2,\sigma^3) \to (-\sigma^1,\sigma^2,\sigma^3)$ in appropriate coordinates. Under $S^3$ parity
\bea
(A_{\mu1},A_{\mu2},A_{\mu3}) &\to& (-A_{\mu1},A_{\mu2},A_{\mu3}) \nonumber \\
(s_\mu^1,s_\mu^2,s_\mu^3) &\to& (-s_\mu^1,s_\mu^2,s_\mu^3)\nonumber\\
e &\to& -e \nonumber\\
\epsilon^{ijk} &\to& - \epsilon^{ijk}\,\nonumber
\eea
therefore
\bea
e ds^i\wedge A_i &\to& - e ds^i\wedge A_i\nonumber \\
e\: \epsilon_{ijk} s^i\wedge s^j \wedge s^k &\to& -e \epsilon_{ijk} s^i\wedge s^j \wedge s^k\,, \nonumber
\eea
cancelling the sign change from spacetime parity.

This Lagrangian is Chern-Simons-like in that
\be
L_{\sdiffthree}=2 \oint d^3\sigma e\,  \delta A_i \wedge F^i -
d \left[ \oint \! d^3\sigma\, e s^i \wedge \delta A_i\right] \, .
\ee
with $F^i$ the $\sdiffthree$ field strength spacetime 2-form:
\be
F^i_{\mu\nu}\equiv \partial_{[\mu} s^i_{\nu]} + s^j_{[\mu} \partial_{|j|}s^i_{\nu]}\,.
\ee
On-shell field configurations are therefore pure gauge, at least locally. This is an ``exotic'' gauge theory (in the terminology of \cite{Bandos:2008jv}) as far as the Lie algebra of $\sdiffthree$ is concerned, but perfectly ordinary from where the Nambu 3-algebra of functions with the bracket (\ref{nambubracket}) is sitting: The spacetime 4-form $d L_{\sdiffthree}$ is the obvious generalisation of the invariant Chern-Weil polynomial to the 3-algebra case. We refer to 15.2.3 of \cite{deAzcarraga:2010mr} for details on this point. 

A less compact---but more useful---form of the $\sdiffthree$ NCS Lagrangian can be obtained by expanding the gauge field $s^i_\mu$ in harmonics. The appropriate set of $S^3$ harmonics is somewhat obscure but has been considered previously in the literature; it is the set of {\it vector hyperspherical harmonics} \cite{Cutkosky:1983jd} which we will denote by ${\cV}^i_\alpha$, following the conventions of \cite{Aharony:2005bq} (but see \cite{Ishiki:2006rt} for a systematic derivation from representation theory). Let us expand
\be
\label{gaugefieldmodeexpansion}
s_\mu^i(x,\sigma) = \sum_\alpha s_{\alpha\mu}(x) \cV_\alpha^i(\sigma)\,,
\ee
The index $\alpha$ labels a state in a particular class of representations of $\mathfrak{so}(4)\cong \mathfrak{su}(2) \oplus \mathfrak{su}(2)$; it denotes the triple of indices
\be
\alpha=(j_\alpha M_\alpha \epsilon_\alpha)
\ee
where $j_\alpha \geq 1$ is an integer labelling a representation of $SO(3) \subset SU(2)\times SU(2) / \mathbb{Z}_2$ and $\epsilon_\alpha=\pm 1$ distinguishes between the ``left'' and ``right'' $\mathfrak{su}(2)$ in $\mathfrak{so}(4) \cong \mathfrak{su}(2) \oplus \mathfrak{su}(2)$, while $M_\alpha=(m_\alpha,\tilde m_\alpha)$ denotes a particular state in the irreducible representation of $\mathfrak{su}(2)\oplus \mathfrak{su}(2)$ specified by $j_\alpha$ and $\epsilon_\alpha$, see appendix \ref{s3harmonics} for more details. 

Now since $d$ acts on $x$ alone the $\sdiffthree$ NCS Lagrangian in harmonics is\footnote{We have solved for the mode expansion of $A_i$ here; this is trivial since $s^i$ and $A_i$ are related by the $S^3$ curl operator which diagonal in the basis we are using.}
\be
\label{csharmonics}
L_{\sdiffthree}=
\sum_\alpha ds_\alpha \wedge (s_{\alpha})^* \left(\frac{-\epsilon_\alpha}{j_\alpha+1}\right)  -\frac{1}{3}\sum_{\alpha \beta \gamma}s_\alpha\wedge s_\beta \wedge s_\gamma E^{\alpha\beta\gamma}\,
\ee
and the gauge transformation of the gauge field mode $s_\alpha$ is:
\be
\label{gaugetransformationharmonics}
\delta_\xi s_\gamma = d \xi_\gamma - \epsilon_\gamma (j_\gamma +1) \sum_{\alpha \beta} \xi_\alpha \wedge s_\beta E^{\bar \gamma \alpha \beta}\,.
\ee
$E^{\alpha \beta \gamma}$ are the structure constants of $\sdiffthree$ and are defined in appendix \ref{s3harmonics}. In this notation, an $S^3$ parity transformation swaps left and right $\mathfrak{su}(2)$ factors \cite{Cutkosky:1983jd}:
\be
\cV_{j_\alpha M_\alpha \epsilon_\alpha} \to (-1)^{j_\alpha+1}\cV_{j_\alpha (M_\alpha)^T (-\epsilon_\alpha)}
\ee

\subsection{Truncation to $\mathfrak{su}(2) \oplus \mathfrak{su}(2)$. Quantization of the NCS level}
\label{sect_truncation}
We will now exhibit the ordinary $\mathfrak{su}(2) \oplus \mathfrak{su}(2)$ CS Lagrangian inside the $\sdiffthree$ NCS Lagrangian. We will first use an abstract argument to see this. Consider a finite number of divergence-free vector fields ${\cal U}_\alpha$ that form a (finite dimensional) Lie subalgebra $K$ of $\sdiffthree$ under the Lie bracket, with structure constants $C^\gamma_{\ \alpha \beta}$:
\be
\label{closedsubalgebra}
[{\cal U}_\alpha,{\cal U}_\beta]^i = C^\gamma_{\ \alpha \beta} ( {\cal U}^i_\gamma)\,.
\ee
If we now write down the vector field ${\cal U}_\gamma$ in terms of the curl of a 1-form ${\mathcal W}_\gamma$, or
\be
({\cal U}^i_\gamma) = \epsilon^{i j k} \partial_j {\mathcal W}_{k\gamma} \qquad \text{where} \qquad \epsilon^{i j k } \equiv e^{-1} \varepsilon^{i j k}
\ee
and given that
\be
{\cal L}_{\cal U} \epsilon^{i j k} =0
\ee
for all divergence-free vector fields ${\cal U}$, we get
\be
{\cal L}_{{\cal U}_\alpha} {\cal U}^i_\beta=[{\cal U}_\alpha,{\cal U}_\beta]^i= \epsilon^{i j k}\partial_j \left({\cal L}_{{\cal U}_\alpha}{\mathcal W}_\beta\right)_k
\ee
using Cartan's magic formula. Setting this equal to (\ref{closedsubalgebra}) implies
\be
({\cal L}_{{\cal U}_\alpha}{\mathcal W}_\beta)_i= C^\gamma_{\ \alpha \beta} {\mathcal W}_{i \gamma} + \partial_i \kappa_{\alpha \beta}\,.
\ee
for some $S^3$ scalar $\kappa_{\alpha \beta}$, which will drop out in the following.

This relation is necessary for proving gauge-invariance (under the subalgebra $K$) of the truncated $\sdiffthree$ CS action. The infinitesimal gauge transformation of $A_i(x,\sigma)$ is
\be
\delta_\xi A_i= d \omega_i -\cL_\xi A_i
\ee
If we use the {\it truncated} field $A_i(x,\sigma)|_K\equiv A^\alpha(x) \cW_{i \alpha}(\sigma)$ then the infinitesimal variation with gauge parameter $\xi^i|_K = \xi^\alpha {\cal U}_\alpha$ closes (up to a total derivative which later drops out):
\be
\delta_{\xi|_K} (A_i)|_K=(dA^\alpha - C^\alpha_{\ \beta\gamma} \xi^\beta A^\gamma) \cW_{i \alpha}\,.
\ee
The gauge variation of $s^i|_K$ also closes (this is trivial since we assumed $K$ is a Lie subalgebra of vector fields).

Let us now split
\be
\mathfrak{sdiff}(S^3)= K+ M
\ee
where $K$ is the (finite dimensional) subalgebra of vector fields $\xi|_K$ and $M$ consists of the rest of the $\sdiffthree$ generators. We truncate by setting the $M$ components of the gauge fields to zero. The gauge invariance of the truncated Chern-Simons type action then follows from
\bea
0&=&\delta_{\xi|_K} L_{\sdiffthree} = 2\oint d^3 \sigma e \: \delta_{\xi|_K} A_i \wedge F^i\\
&=& 2\oint d^3 \sigma e \: \left(\delta_{\xi|_K} A_i|_K\right) \wedge F^i|_K + \left(\delta_{\xi|_K} A_i|_M\right) \wedge F^i|_K\,.
\eea
In the second line we set $s|_M=0$ in the field strength 2-form; we have already varied the full action so this is allowed. We have also dropped the total $S^3$ derivatives arising from the variation of $A$ because $F^i$ is divergence-free in its $S^3$ index.

The first term in the above equation is the gauge variation of the truncated action, therefore we need the second term to vanish. A sufficient condition for this is
\be
[M,K] \subseteq M\,.
\ee
This property indeed holds for $K = SO(4) \subset \sdiffthree$. The Lie bracket of two vector harmonics reads
\be
[\cV_\alpha, \cV_\beta] \propto E^{\alpha \beta \gamma} \cV_\gamma\,.
\ee
The coefficient $E^{\alpha \beta \gamma}$ is totally antisymmetric and vanishes whenever $\cV_\alpha, \cV_\beta \in \mathfrak{so}(4), \cV_\gamma \in M$ (see appendix \ref{s3harmonics}).

This demonstrates that the $\sdiffthree$ NCS action reduces to an $SO(4)$ CS action. For our purposes we will need a more explicit formula; thankfully the mode expansion (\ref{gaugefieldmodeexpansion}) is well-adapted to this purpose. In this notation, the truncation to $SO(4)$ is realised by discarding all $j_\alpha >1$. Then $s_{\epsilon=+}$ (resp. $s_{\epsilon=-}$) transforms as a gauge connection in the spin 1 representation of the left (resp. right) $SU(2)$ and is inert under the other factor. The reality condition $s_\mu^i =(s_\mu^i)^\ast$ in conjunction with the mode expansion imply that $s_{1,2,3;+}$ defined as
\begin{align}
s_{\pm 1;\epsilon=+} &\equiv s_{1;+} \pm i s_{2;+} \\
s_{0;\epsilon=+} &\equiv \sqrt{2} i s_{3;+}
\end{align}
(resp. for $s_{\epsilon=-}$) are real. Then an explicit calculation of $E^{\alpha \beta \gamma}$ in (\ref{csharmonics}) gives
\be
\frac{2}{ \pi^2} S_{\sdiffthree; j_\alpha=1} = \left[ \int \tr \left( s_+ \wedge d s_+ + \frac{2}{3} i s_+^3 \right) -\int \tr \left( s_- \wedge d s_- + \frac{2}{3} i s_-^3 \right)\right]
\ee
where 
\be
s_{\pm} = s_{1;\pm} \sigma^1 + s_{2;\pm} \sigma^2 + s_{3;\pm} \sigma^3
\ee
and the $\sigma^i$ are the Pauli matrices. The right-hand side is immediately recognised as the CS action for gauge group $G=SO(4)$ expressed in terms of its left and right $SU(2)$ gauge fields, and is known to be gauge invariant when its prefactor is \cite{Distler:2008mk}
\be
\frac{k}{4 \pi}, \qquad k \in \mathbb{Z}\, .
\ee

We now come to the issue of the quantization of the Chern-Simons level of the $\sdiffthree$ NCS action. Consider acting with a {\it finite} $SO(4)$ gauge transformation on a field configuration where only the $j_\alpha=1$ mode has been switched on. It is clear from formula (\ref{gaugetransformationharmonics}) and the preceding discussion that this transformation will not switch on any modes of $j_\alpha>1$. Therefore the $\sdiffthree$ NCS action inherits the quantization condition of $SO(4)$ CS theory. By the discussion above, the $\sdiffthree$ action is properly normalised as
\be
\label{normalisedsdiff}
\frac{k}{2\pi^3}S_{\sdiffthree}
\ee
and $k \in \mathbb{Z}$ is a {\it necessary} condition for invariance under finite $SO(4)$ gauge transformations. Let us phrase this as a
\begin{proposition}
Let $S_{\sdiffthree}$ be the spacetime integral (\ref{sdiffaction}) of the $\sdiffthree$ NCS Lagrangian 3-form as defined previously (with an $S^3$ of radius unity). Under a finite gauge transformation with gauge parameter $g_{SO(4)}$ taking values in $SO(4) \subseteq \sdiffthree$, we have
\be
\frac{k}{2\pi^3}\left(S_{\sdiffthree}[g_{SO(4)}A|_{SO(4)}]-S_{\sdiffthree}[A|_{SO(4)}]\right) =0  \mod 2\pi\mathbb{Z} \iff k \in \mathbb{Z}
\ee
where $g_{SO(4)}A|_{SO(4)}$ is the transformed gauge field $A|_{SO(4)}$ entering the $S_{\sdiffthree}$ action, and $|_{SO(4)}$ indicates that only the $SO(4)$ modes are nonvanishing.
\end{proposition}
We stress again that we have only shown a necessary condition for gauge invariance under finite $\sdiffthree$ transformations. In light of the fact that $SO(4) \subset \sdiffthree$ is a homotopy equivalence (proven in appendix \ref{appendix_instantons}) one naturally expects that this is also sufficient, but properly speaking one would have to study $\sdiffthree$ NCS theory in the manner of Dijkgraaf and Witten \cite{Dijkgraaf:1989pz} to prove this. 

\subsection{Interpretation as higher spin gravity}
\label{sect_higherspin}
For convenience, we reproduce the mode expansion of the $\sdiffthree$ NCS action from before
\be
L_{\sdiffthree}=
\sum_\alpha ds_\alpha \wedge (s_{\alpha})^* \left(\frac{-\epsilon_\alpha}{j_\alpha+1}\right)  -\frac{1}{3}\sum_{\alpha \beta \gamma}s_\alpha\wedge s_\beta \wedge s_\gamma E^{\alpha\beta\gamma}\,,
\ee
and also the gauge transformation of the gauge field mode $s_\alpha$:
\be
\delta_\xi s_\gamma = d \xi_\gamma - \epsilon_\gamma (j_\gamma +1) \sum_{\alpha \beta} \xi_\alpha \wedge s_\beta E^{\bar \gamma \alpha \beta}
\ee
Could this describe a (Euclidean) theory of higher spin de Sitter gravity? The lowest $\mathfrak{su}(2)$ spin ($j_\alpha =1$) sector is certainly identical to ordinary Euclidean Einstein gravity in 3D de Sitter space \cite{Banados:1998tb}. To answer the question, we will follow Blencowe \cite{Blencowe:1988gj} and Campoleoni et al. \cite{Campoleoni:2010zq} and examine whether the linearised theory around de Sitter space has a higher spin interpretation.


Let us produce the full equations of motion:
\be
\label{fullequations}
2 \epsilon_\alpha d (s_\alpha)^\ast + (j_\alpha +1)\sum_{\beta\gamma} s_\beta \wedge s_\gamma E^{\alpha \beta \gamma} =0\,.
\ee
We first need to check that de Sitter space is a solution. Although this follows from the analysis of the preceding subsection, it is easy to check directly. Since we know that the $j_\alpha=1$ action is that of Euclidean de Sitter 3D gravity, we need only check that
\be
s_\alpha = \delta_{j_\alpha,1} \bar s_{\alpha} \qquad \text{(no summation)}
\ee
solves the full field equations whenever $\bar s_\alpha$ solves the truncated equations. There are two cases to consider:
\begin{itemize}
\item $j_\alpha=1$: We know that $E^{\alpha \beta \gamma}$ is only nonzero when $j_\beta=j_\gamma$ (see discussion around (\ref{f})). The $j_\beta>1$ contributions vanish; what remains is the {\it truncated} field equation which $\bar{s}_\alpha$ solves by assumption.
\item $j_\alpha >1$: The potentially problematic terms are the $j_\beta=1$ and/or $j_\gamma=1$ ones. By the previous remark and antisymmetry of $E^{\alpha \beta \gamma}$ only one of $j_\beta,j_\gamma$ can equal 1, but in this case (\ref{fullequations}) reduces to $0=0$.
\end{itemize}

It remains to show that the action expanded to quadratic order around this background actually describes higher spin fields. Let us write 
\be
s_\alpha = \delta_{j_\alpha,1} \bar s_{\alpha} +h_\alpha
\ee
where $h_\alpha$ is the perturbation. The quadratic action is (for the $\sdiffthree$ constants $f^\gamma_{\alpha \beta}$ see (\ref{f}) in appendix \ref{s3harmonics}):
\be
\label{freesdiffaction}
S_{\rm quad}=\int \left[\left(-\frac{\epsilon_\alpha}{j_\alpha+1}\right) \sum_\alpha dh_\alpha \wedge (h_\alpha)^\ast -\sum_{\alpha\beta\gamma} h_\alpha\wedge h_\beta \wedge \bar{s}_\gamma f^\gamma_{\alpha\beta}\right]\,.
\ee
As one might have already guessed this action turns out to describe higher spins in a frame-like formulation, so we will give a very brief description of this formalism here, following \cite{Campoleoni:2010zq}. For a spin $s\geq 2$ field one has a generalised dreibein $e_\mu^{a_1 a_2 \dots a_{s-1}}$ and a generalised spin connection $\omega_\mu^{a_1 a_2 \dots a_{s-1}}$, both symmetrised and traceless in their flat indices. These indices describe the transformation properties of the fields under local Lorentz transformations. In the context of Euclidean de Sitter gravity, these are the gauge transformations associated to the diagonal $SO(3)$ in $SU(2)\times SU(2)/\mathbb{Z}_2$, which is the ``Lorentz'' group in this case. $\omega$ and $e$ have the \emph{intrinsic} parity assignments
\bea
e^{a_1 a_2 \dots a_{s-1}}&\to& -e^{a_1 a_2 \dots a_{s-1}}\\
\omega^{a_1 a_2 \dots a_{s-1}} &\to& +\omega^{a_1 a_2 \dots a_{s-1}}
\eea
These can be easily understood from the CS point of view. The generalised dreibein and spin connection are packaged into two independent $SU(s)$ gauge fields $A_\pm$ as
\be
A_\pm=\omega \pm e
\ee
and the higher spin action takes the form
\be
S_{\rm CS}[A_+]-S_{\rm CS}[A_-]\,.
\ee
This is invariant under parity if we assign the parity transformation $A_+ \leftrightarrow A_-$, implying the previous $e$ and $\omega$ parity assignments.

A mode $h_\alpha$ of the $\sdiffthree$ gauge field is not irreducible under the diagonal $SO(3)$, but transforms as (see appendix \ref{s3harmonics})
\be
\label{diagonalso3decomposition}
\frac{j_\alpha+1}{2} \otimes \frac{j_\alpha-1}{2} = \mathbf{1} \oplus \mathbf{2} \oplus \dots \oplus j_\alpha, \qquad (\mathbb{Z} \ni j_\alpha\geq 1)\,,
\ee
suggesting that the $\sdiffthree$ NCS action propagates a countable number of massless fields of all integer spins $s\geq 2$ around its Euclidean de Sitter vacuum. There are two complications here however: The first is that the $\sdiffthree$ NCS action clearly propagates zero degrees of freedom on-shell and it is not a priori clear what spin means in this context. This problem also comes up when we consider $\mathfrak{sl}(n)\oplus \mathfrak{sl}(n)$ CS higher spin gravity, and was resolved by Blencowe; the upshot of the discussion in \cite{Blencowe:1988gj} is that the 3D linearised equations of motion are matched to the dimensionally-reduced equations of free 4D higher spin fields.

The other complication is that while expressions for the $\sdiffthree$ constants $f^\gamma_{\alpha \beta}$ certainly exist (in appendix \ref{s3harmonics} of this paper), they are in terms of Clebsch-Gordan coefficients and we could not work with them in generality. This can however be circumvented. The quadratic action (\ref{freesdiffaction}) has a very particular Chern-Simons-like form, and we will show that the equations of motion derived from this action are fixed by the twin requirements of linearised ``higher spin'' gauge invariance (of the form $\delta h_\alpha = d\xi_\alpha +\dots$) and \emph{background} gauge invariance under the diagonal $SO(3)$, under which both the field perturbation $h_\alpha$ and the background gauge field $\bar s_\alpha$ transform. Once this uniqueness is shown, one need only show that the resulting equations indeed describe higher spin gauge fields in the sense of Blencowe above; thankfully this has been thoroughly demonstrated by Campoleoni et al. for the $AdS_3$ 
case \cite{Campoleoni:2010zq} and it also follows from the observation that $\mathfrak{su}(n)\oplus \mathfrak{su}(n)$ CS theory describes Euclidean de Sitter higher spins \cite{Basu:2015exa}.

Consider a pair of 1-form field perturbations
\be
(H^A_e,H^A_\omega)=H^A
\ee
which have the same intrinsic parity assignments as $e$ and $\omega$ and are valued in an arbitrary irreducible representation of the diagonal $SO(3)$, and consider also the background dreibein and spin connection which we will collectively denote by $E^a$. We are using different indices to stress that $H^A$ and $E^a$ transform differently under local $SO(3)$. Invariance under local $SO(3)$ corresponds to \emph{background gauge transformations}, under which $H^A$ varies tensorially, but some components of $E^a$ (specifically, the spin connection) do not.

Let us assume that $H^A$ has an ordinary gauge invariance of the form
\be
\label{freehsgaugevariation}
\delta_\Xi H^A= d \Xi^A + \mathbf{N}^A_{a B} E^a \Xi^B
\ee
where $\mathbf{N}^A_{a B}$ is a spacetime-constant $SO(3)$ tensor and $\Xi^A$ is a 0-form gauge parameter valued in the same representation as $H^A$, and that $H^A$ satisfies a field equation
\be
\label{freehsfieldequation}
0= F(H)^A \equiv d H^A + \mathbf{M}^A_{b B} E^a \wedge H^B
\ee
where $\mathbf{M}^A_{a B}$ is another constant $SO(3)$ tensor. Gauge invariance of the field equation under $\Xi^A$-gauge transformations implies
\be
0=(\mathbf{M}^A_{b B}-\mathbf{N}^A_{a B}) E^a \wedge d\Xi^B+ (\mathbf{N}^A_{a C} dE^a  + \mathbf{M}^A_{a B}\mathbf{N}^B_{b C} E^a \wedge  E^b )\Xi^C
\ee
therefore
\be
\mathbf{M}^A_{b B}=\mathbf{N}^A_{b B}
\ee
and
\be
\label{freehscond2}
0=\mathbf{N}^A_{a C} dE^a  + \mathbf{N}^A_{a B}\mathbf{N}^B_{b C} E^a \wedge  E^b
\ee
are necessary and sufficient conditions for gauge invariance. The first equation fixes the field equation in terms of the gauge transformation, while the second is a condition on the background field alone, and should be equivalent to---or be implied by---the equation of motion for the background fields. Note that this second equation takes the same form as the field equations in the dreibein formalism:
\bea
0&=&de^a + \epsilon^{abc} \omega_b \wedge e_c \\
\epsilon^{abc} e_b \wedge e_c &\propto& d\omega^a + \frac{1}{2} \epsilon^{abc} \omega_b\wedge \omega_c
\eea
where the proportionality constant in the last formula is of course the cosmological constant.

Therefore to specify the field equation for free higher spins we need three things:
\begin{itemize}
\item An on-shell background dreibein and spin connection $E^a$;
\item a representation of the local ``Lorentz'' group $SO(3)$ under which the 1-form field $H^A$ as well as the background $E^a$ transforms;
\item and an $SO(3)$ tensor $\mathbf{N}^A_{b B}$\,.
\end{itemize}
Local $SO(3)$ invariance actually fixes the tensor $\mathbf{N}^A_{b B}$; since the field equation must be invariant under $SO(3)$ gauge rotations that act on both $H^A$ and $E^a$ the field equation can be rewritten as
\be
0= \bar DH^A + \mathbf{T}^A_{bB} E^b \wedge H^B
\ee
where $\bar D$ is now the background exterior $SO(3)$-covariant derivative and $\mathbf{T}$ is an \emph{invariant} tensor of $SO(3)$. $\mathbf{T}$ can be further specified as follows: $E^b \wedge H^B$ transforms in the $\mathbf{1} \otimes j$ representation of $SO(3)$ (where $j+1$ is the spin $j+1\geq 2$ of the free higher spin field $H$), which decomposes as
\be
\mathbf{1} \otimes j= (j+1) \oplus j \oplus (j-1)
\ee
Since $\bar D H^A$ by definition transforms in the $j$ representation of $SO(3)$, $\mathbf{T}$ is fixed up to a constant (itself fixed by (\ref{freehscond2})) by the requirement that it projects $E^b \wedge H^B$ onto its $j$ subrepresentation which only appears once in the sum, and is therefore some Clebsch-Gordan coefficient. To be more concrete, write
\be
H^A= (H_e^{(a_1 a_2\dots a_{j})},H_\omega^{(a_1 a_2\dots a_{j})})\
\ee
where each index $a_i$ is in the $1$ of $SO(3)$ and we are taking the symmetric and traceless part; then the only $SO(3)$ covariant expressions linear in $E^b \wedge H^B$ transforming in the same representation as $H$ are
\be
\mathbf{T}^A_{bB} E^b \wedge H^B=\epsilon^{cd(a_1} e_c\wedge H_{e,\omega \: d}^{\phantom{e,\omega \: d} a_2\dots a_{j})}\,,
\ee
and the only possible \emph{parity-preserving} field equations look like
\begin{align}
\label{freehsequation_dreibein}
0&=\bar D H_e^{a_1 a_2 \dots a_j} + \epsilon^{cd(a_1} e_c\wedge H_{\omega \: d}^{\phantom{\omega \: d} a_2\dots a_{j})}\, \\
\label{freehsequation_spin}
0&=\bar D H_\omega^{a_1 a_2 \dots a_j} + \epsilon^{cd(a_1} e_c\wedge H_{e \: d}^{\phantom{e \: d} a_2\dots a_{j})}
\end{align}
where we have ignored the relative constants. These expressions match (2.28) of \cite{Campoleoni:2010zq} (with $ H_e\to h$ and $H_\omega \to v$). We have just proven the
\begin{proposition}
Let $H_\omega^A$ and $H_e^A$ be spacetime 1-form fields in three dimensions valued in the same representation of $SO(3)$ (as indicated by the index $^A$), and with the intrinsic parity assignment
\be
H_\omega^A \to + H_\omega ^A, \quad H_e^A \to - H_e^A\,.
\ee
Let also $E^A$ collectively denote a background dreibein $e^a$ and spin connection $\omega^a$, which are also spacetime 1-forms, parity eigenstates, and are valued in the $\mathbf{1}$ of $SO(3)$, with $\omega^a$ transforming as an $SO(3)$ connection and $e^a$ transforming tensorially, which are also on-shell in the sense that they satisfy a (parity-preserving) equation of the form (\ref{freehscond2}). If $H_{e,\omega}^A$ satisfy an $SO(3)$- and parity-covariant linear equation of motion of the form (\ref{freehsfieldequation}), invariant under a gauge transformation of the form (\ref{freehsgaugevariation}), then the form of (\ref{freehsfieldequation}) is uniquely specified once (\ref{freehscond2}) and the transformation (\ref{freehsgaugevariation}) are fixed.
\end{proposition}
This proposition was hinted at in \cite{Campoleoni:2010zq}.

Let us see how this applies to the quadratic action around the de Sitter vacuum of $\sdiffthree$ gauge theory. The field equation derived from the action (\ref{freesdiffaction}) is of the form (\ref{freehsfieldequation}) discussed above, as is the linearised gauge variation. The analysis in $SO(4)$ harmonics implies, as outlined above, that an irreducible diagonal-$SO(3)$ multiplet may be specified by a pair of integers $(J, j)$ where $J \geq 1$ is equal to the integer $j_\alpha$ specifying an $SO(4)$ irrep as before and $1\leq j \leq J$ labels a representation in the decomposition (\ref{diagonalso3decomposition}) under the diagonal $SO(3)$ of the corresponding mode $h_\alpha$ of the field perturbation. Now \emph{suppose we keep just the $(J, j)$ harmonics}, setting all other $h_\alpha$ to zero. This is an $SO(3)$ and parity invariant condition which breaks all linearised higher spin gauge invariances apart from those of the form (\ref{freehsgaugevariation}), i.e. with a gauge parameter $\Xi_\alpha$ 
restricted 
to the $(J,j)$ harmonic. From the preceding argument, the linearised equations of motion for the $(J,j)$ harmonic mode should therefore be identical to the free higher spin equations of motion for spin $j+1$, i.e. (\ref{freehsequation_dreibein}) and (\ref{freehsequation_spin}) once we combine the $h_\alpha$ into parity eigenstates. This gives the pyramid of spins (\ref{pyramid_of_spins}) in the Introduction, where each line is specified $J=j_\alpha$ has maximum spin $j_\alpha+1$.

We have verified that this is true for the $(2,2)$ harmonic by comparison to the (linearised) spin 3 action of Campoleoni et. al \cite{Campoleoni:2010zq}, where also the equivalence of this frame-like formalism to the one of Fronsdal is established. This is a straightforward but tedious and somewhat lengthy calculation which we will only summarise here. The relevant field perturbations $h_\alpha$ are given in the $(\epsilon_\alpha=+) \leftrightarrow \mathbf{1/2}\otimes \mathbf{3/2}$ and $ (\epsilon_\alpha=-)\leftrightarrow \mathbf{3/2}\otimes \mathbf{1/2}$ representations of $SU(2)\times SU(2)/\mathbb{Z}_2$, both of which decompose as
\be
\mathbf{2} \oplus \mathbf{1}
\ee
under $SO(3)$. The modes in the $\mathbf{1}$ multiplet are superfluous and are set to zero. On the other hand, the generalised dreibein and spin connection in the usual formulation of free higher spins in the frame-like formalism are symmetric traceless 2-index tensors of $SO(3)$, which means they appear as the $\mathbf{2}$ in the decomposition
\be
\mathbf{1}\otimes \mathbf{1}= \mathbf{2}\oplus \mathbf{1} \oplus \mathbf{0}\,.
\ee
The two actions can be compared once both changes of basis $( \mathbf{3/2}\otimes  \mathbf{1/2} \to \mathbf{2})$ and $(\mathbf{1}\otimes \mathbf{1} \to \mathbf{2})$ have been performed and the relevant $\sdiffthree$ constants $f^\alpha_{\beta \gamma}$ have been tabulated; we performed this with the help of Mathematica routines.

\section{The ${\rm SDiff}(S^3)$ BLG model}
The BLG action describes a non-abelian $N=8$ supermultiplet of
\begin{itemize}
\item Scalar fields $\phi^I(x,\sigma) \:(I=1,\dots 8)$ in the $\mathbf{8}_{\rm V}$ representation of the R-symmetry group $Spin(8)$
\item Majorana anticommuting $SL(2;\mathbb{R})$ spinor fields $\psi_A(x,\sigma) \: (A=1,\dots 8)$ in $\mathbf{8}_{\rm S}$
\item Gauge fields $A_{\mu i}(x,\sigma)$ (as before)
\end{itemize}
where all matter fields are $G=\sdiffthree$ scalar fields under $\sdiffthree$ gauge transformations, i.e.
\be
\delta_\xi \phi^I = -\xi^i \partial_i \phi^I
\ee
and the $\sdiffthree$ covariant derivative is
\be
{\cal D}_\mu \phi^I\equiv \partial_\mu \phi^I + s_\mu^{\ i} \partial_i \phi^I\,.
\ee
The Lagrangian density can be written (suppressing $Spin(8)$ indices where appropriate)
\begin{align}
L_{\rm BLG}= \oint d^3 \sigma\: e \left[ -\frac{1}{2} | {\cal D}\phi|^2 - \frac{i}{2} \bar{\psi} \gamma^\mu {\cal D}_\mu \psi + \frac{i}{4} \epsilon^{i j k}\partial_i\phi^I \partial_j \phi^J \left(\partial_k \bar{\psi} \rho^{IJ} \psi\right)  \right. \nonumber \\
\left.- \frac{1}{12} \{\phi^I, \phi^J, \phi^K \}^2 \right]+\frac{1}{2} L_{\sdiffthree}\,.
\end{align}
where $ \{\phi^I, \phi^J, \phi^K \}$ is the Nambu 3-bracket:
\be
 \{\phi^I, \phi^J, \phi^K \} \equiv \epsilon^{ijk} \partial_i \phi^I \partial_j \phi^J \partial_k \phi^K
\ee
Details on $\gamma^\mu$ and $\rho^{IJ}$ matrices, as well as proof that this Lagrangian enjoys $OSp(8|4)$ superconformal symmetry can be found in the reference \cite{Bandos:2008jv}. See also \cite{Cederwall:2008vd} for a reformulation with manifest supersymmetry.

This theory has a $1/2$ BPS smooth funnel solution describing M2 branes blowing up into a planar M5: Consider the configuration
\be
\label{funnel}
\phi^a= \frac{1}{\sqrt{2x^1}} X^a(\sigma), \quad a=1,2,3,4\,,
\ee
where the $X^a(\sigma)$ parameterise a unit 3-sphere as
\be
\sum_{a=1}^4 (X^a(\sigma))^2=1,\quad \{X^a,X^b,X^c\} = \epsilon^{abc}_{\phantom{abc}d} X^d\,,
\ee
and all other fields are set to zero. This configuration describes a planar M5 brane at $x^1=0$ with an M2 brane ``spike'' of 3-sphere cross-section. As shown in \cite{Bandos:2008jv}, this solves the BPS equation
\be
\rho^I \gamma^\mu \cD_\mu\phi^I \epsilon = \frac{1}{6} \{ \phi^J,\phi^K, \phi^L\} \rho^{JKL} \epsilon\,.
\ee
This is the Basu-Harvey equation \cite{Basu:2004ed}, which arises in all BLG models as a BPS condition. When the BLG gauge group $G$ is $SO(4)\subset \sdiffthree$, the corresponding solution (\ref{funnel}) is interpreted as a \emph{fuzzy} funnel; this parallels the better understood type IIB D1-D3 brane intersections which can be obtained from the M2-M5 intersections we are discussing by compactification followed by T-duality.

There is a single, dimensionless, coupling constant in the $\sdiffthree$ BLG model, which appears in the action as a prefactor, and may be identified with the invariant volume $\oint d^3 \sigma \: e$ of the three-sphere\footnote{To compare with \cite{Bandos:2008jv}: It was pointed out in that paper that one can redefine the dynamical fields such that the coupling $g$ appears only outside the action. Under the simultaneous scaling $e\to \lambda e, A\to \lambda A$ and $(\phi,\psi) \to \sqrt{\lambda} (\phi,\psi)$ the action goes as $S \to \lambda^2 S$, thus the coupling $g$ of \cite{Bandos:2008jv} is identified with the volume {\it squared} of the $S^3$. The sign of the coupling does not appear to make a difference in our analysis so we only consider the positive case.}. Invariance of the $\sdiffthree$ NCS term under finite $SO(4)$ 
gauge transformations forces it to take integer values, and the properly normalised BLG action takes the form
\be
S_{\rm BLG}= \frac{k}{\pi^3} \int L_{\rm BLG}\qquad \text{where} \qquad \oint d^3 \sigma \: e=2 \pi^2,\, k \in \mathbb{Z} \,.
\ee

Let us take a moment to comment. The first thing to notice is that with no continuous dimensionless couplings in the action, the $\sdiffthree$ BLG theory must be conformally invariant to all orders in perturbation theory. The second, more puzzling observation, is that the theory appears to be weakly coupled when the Chern-Simons level $k$ is large, but one would not na\"ively expect any tunable parameters from the M2 brane picture. This was first pointed out for the $\mathfrak{su}(2)\oplus \mathfrak{su}(2)$ BLG Lagrangian (which also has a single integer parameter) by Raammsdonk \cite{VanRaamsdonk:2008ft}. Similar observations can be made for ABJM theory. In either case the resolution involves the relation between the M theory and type IIA pictures.

We will sketch the mechanism originally given by Townsend for the centre of mass modes \cite{Townsend:1995af} or equivalently for a single brane. A D2-brane spans 3 worldvolume dimensions; in a ten-dimensional spacetime its collective coordinates are 7 transverse scalars and a single $U(1)$ gauge field. The gauge field can be dualised to a {\it periodic} scalar corresponding to the position of the brane in the 11th compact dimension; the periodicity comes about because the translation symmetry (properly, shift symmetry) of that scalar is broken by {\it instantons} in the 3D theory \cite{Seiberg:1997ax} (what we would call garden-variety magnetic monopoles for the $U(1)$ gauge field in one dimension higher) whose charge is interpreted as momentum along the 11th compact dimension. The upshot is that the moduli space is given by $\mathbb{R}^7 \times S^1$, where the $S^1$ radius is proportional to the coupling. In the infrared/strong coupling limit that radius blows up and we obtain the moduli space of a single 
M2 brane, which is simply 
$\mathbb{R}^8$. Weak 
coupling would therefore seem to be associated to a compactification limit.

The integer $k$ of BLG/ABJM was ultimately found to be related to the spacetime geometry probed by the branes through analysis of the supersymmetric vacuum moduli space. For $G=U(n)\times U(n)$ ABJM theories at CS level $k$ this is simply \cite{Aharony:2008ug}
\be
{\cal M} = (\mathbb{R}^8/\mathbb{Z}_k)^n / {\cal S}_n\,,
\ee
describing $n$ indistinguishable M2 branes probing $\mathbb{R}^8/\mathbb{Z}_k$. The results for $G=SO(4), SU(2)\times SU(2)$ BLG theories generally fail to have such a straightforward interpretation except for the very lowest values of $k$, but it has been argued that they involve some sort of generalised orbifold. In either case, $k \to \infty$ is a compactification limit, as expected. A crucial role in the analyses of \cite{Lambert:2008et,Distler:2008mk,Lambert:2010ji,Bagger:2012jb,Aharony:2008ug} is played by $G$-instantons, which break what would na\"ively be a continuous $U(1)$ identification of $\mathbb{R}^8$ \cite{VanRaamsdonk:2008ft} down to a discrete $\mathbb{Z}_k$ subgroup; only the latter quotient makes any sense.

In the next section we will perform a straightforward calculation of the vacuum moduli space, which---in contrast to the $SO(4)$ model---turns out to have a nice interpretation. This is rather fortuitous since the more sophisticated approach of \cite{Lambert:2008et,Distler:2008mk,Lambert:2010ji} (which gives the correct result for $G=SO(4), SU(2)\times SU(2)$) appears to fail for $G=\sdiffthree$. Unexpectedly, this is not due to $\sdiffthree$-instantons, which turn out to be very well behaved.

\subsection{Vacuum moduli space}
\label{sect_vacuummoduli}
The classical vacuum moduli space of the $\sdiffthree$ BLG theory of \cite{Bandos:2008jv} is defined to be the space of all on-shell bosonic field configurations that preserve all 16 supersymmetries, {\it modulo} gauge transformations. Inspection of the supersymmetry transformations (formula (4.4) of \cite{Bandos:2008jv}) shows that this is equivalent to
\be
s_\mu^i=\psi=\partial_\mu\phi^I=\{\phi^I,\phi^J,\phi^K\}=0\qquad \forall   I=1,2,\dots 8
\ee
If we rephrase a bit, the last equality states that the maximum rank of the Jacobian matrix of the mapping
\be
\vec{\phi}: S^3 \to \mathbb{R}^8
\ee
is two. Therefore, a {\it generic} point in moduli space is given by a mapping into what is {\it locally} a 2-manifold $M_2$.
\be
\vec{\phi}: S^3 \to M_2 \subseteq \mathbb{R}^8\,.
\ee
Making the further assumption that $M_2$ is a manifold {\it globally} it is clear that $M_2$ is actually a 2-sphere\footnote{This should not be restrictive: The codimension is high enough that one can remove any self-intersections by pushing coincident points slighly apart in the transverse directions, and any cusp-like singularities can presumably be obtained as limits of smooth spherical embeddings. In any case we will simply \emph{assume} that the image is an embedded $S^2$.} and these generic points in moduli space describe embeddings of the Hopf bundle
\be
S^1\to S^3 \xrightarrow{f} S^2
\ee
in $\mathbb{R}^8$.

Maps $f: S^3 \to S^2$ are classified up to homotopy by the {\it Hopf invariant} (see e.g. the textbook \cite{Monastyrsky:1993ig}), which we will define. Let $\Omega$ be a volume 2-form on $S^2$ and consider its pullback $f^\ast \Omega$ to $S^3$. While $\Omega$ is by definition a closed but not exact 2-form on $S^2$, its pullback to $S^3$ is necessarily exact (the relevant de Rham cohomology for $S^3$ is trivial). We can therefore define a 1-form $\omega$ on $S^3$ by
\be
d \omega \equiv f^\ast \Omega
\ee
and the Hopf invariant is given by
\be
I(f)\equiv \int_{S^3} \: \omega \wedge d\omega\,.
\ee
The appropriately normalised Hopf invariant is always an integer. Hopf's original map $h: S^3 \to S^2$ has Hopf invariant equal to 1. One can compose $h$ with a map $w_k: S^3 \to S^3$ that winds $k \in \mathbb{Z}$ times to obtain a map of any Hopf invariant $k$, i.e. a representative in each homotopy class. Since a map $w_k, k\neq 1$ fails to be volume-preserving, it does {\it not} lie in ${\rm SDiff}(S^3)$; hence \emph{the Hopf invariant also distinguishes between different branches of our moduli space}.

The upshot of the discussion so far is that generic points in $\sdiffthree$ BLG moduli space are specified by
\be
\label{pre_moduli_space}
I\in \mathbb{Z}^*\,, \qquad \iota: S^2 \to \mathbb{R}^8\,,
\ee
i.e. the non-vanishing Hopf invariant\footnote{If the Hopf invariant vanishes the $S^2$ will generally fail to be an embedded submanifold. The reason is that the map $S^3 \to S^2$ is null-homotopic in this case and therefore generally fails to be onto.} and the embedding $\iota$ into $\mathbb{R}^8$. However, we have yet to quotient by the gauge group. If we let $(z,\theta), z \in (\mathbb{C}\cup \infty) \cong S^2, \theta \in S^1$ parameterise a local trivialisation of the Hopf fibration, then we can partially gauge fix $\sdiffthree$ by setting
\be
\partial_\theta \phi^I=0 \iff \phi^I=\phi^I(z)\,.
\ee
There is a residual gauge group of fibre-preserving---in the sense that they map fibres to fibres---diffeomorphisms inside $\sdiffthree$, which \emph{a priori} fill out a subgroup of ${\rm Diff}(S^2)$. If we remove the volume-preservation restriction, it is certainly true that fibre-preserving transformations project down to the full diffeomorphism group of the $S^2$ base (this is the content of the ``Projection theorem'' of Cerf and Palais, see e.g. \cite{Hong}). Whether this is still true if we restrict to fibre-preserving, total-space-volume-preserving diffeomorphisms is apparently unknown for nontrivial bundles but is clearly true for trivial ones. Assuming that this is true, the manifold of supersymmetric vacua is simply
\be
\label{moduli_space}
{\cal M}(\sdiffthree \text{ BLG})=\frac{{\rm Embeddings}(S^2 \to \mathbb{R}^8)}{{\rm Diff}(S^2)} \times \mathbb{Z}
\ee
with the proviso that ${\rm Diff}(S^2)$ denotes the identity component; this avoids double counting. We have included the vanishing Hopf invariant sector here with the understanding that it contains all the \emph{non}generic points of moduli space, which are not embeddings of $S^2$. This moduli space can be interpreted as the space of configurations of a 2-sphere's worth of indistinguishable M2 branes, each fully localised on $\mathbb{R}^8$.

Since ${\cal M}(\sdiffthree \text{ BLG})$ describes embeddings of (the base of) the Hopf fibration in the transverse coordinates, it seems tempting to interpret the vacuum moduli spaces of $G=SU(2)\times SU(2)(/\mathbb{Z}_2)$ BLG theories as \emph{fuzzy Hopf fibrations}, mirroring the interpretation of the $1/2$ BPS funnel solutions of the previous section as ``fuzzy funnels''. In particular, a point on the moduli space should represent a fuzzy 2-sphere embedded in $\mathbb{R}^8$, and the fuzzy Hopf fibration should then be understood as the map from an abstract fuzzy $S^3$ (which has a transitive action of the gauge group $G$) to the fuzzy 2-sphere.

Let us compare with $G=SU(2)\times SU(2)(/\mathbb{Z}_2)$ BLG. The corresponding moduli space can be obtained from the above discussion by replacing every instance of a diffeomorphism group with the corresponding isometry group, and the 2-sphere by the pair of points $z \in \mathbb{C}$; the result is $(\mathbb{R}^8)^2/O(2)$ \cite{VanRaamsdonk:2008ft}. This conflicts with expectations from supersymmetry---at the very least we would expect a manifold of even dimensionality---and also fails to admit an interpretation as the configuration space for some number of indistinguishable M2 branes fully localised on the transverse $\mathbb{R}^8$. The more sophisticated calculations of references \cite{Lambert:2008et,Distler:2008mk,Lambert:2010ji} instead give $(\mathbb{R}^8)^2/\Gamma_k$, where $\Gamma_k$ is a dihedral group whose order depends on $k$ and on whether we include the $\mathbb{Z}_2$ quotient in the gauge group $G$. The same methods also give the correct result for ABJM theory \cite{Aharony:2008ug}. The 
essential step appears to involve carefully integrating out the unbroken gauge field $A_H$ (where the abelian $H\subset G$ acts trivially on scalars in $\cal M$) in the low-energy effective action describing motion in the moduli space. This is complicated because the gauge field associated to the \emph{broken}, residual gauge group $K$ (where $K=O(2)$ for $G=SO(4)$ or $SU(2)\times SU(2)$ and $K={\rm Diff}(S^2)$ for $G=\sdiffthree$) couples to it through a $BF$ type term:
\be
\int A_H \wedge F_{K}\,.
\ee
When $K=O(2)$, $F_K$ is an abelian field strength. Integrating by parts produces an $F_H=dA_H$ term which can be dualised into a scalar; this is a \emph{periodic} scalar whose periodicity is controlled by
\be
\pi_2(G/H)
\ee
and is ultimately responsible for breaking $K=O(2)$ down to $\Gamma_k$.

When $K={\rm Diff}(S^2)$ ($G=\sdiffthree$), the field strength $F_K$ is nonabelian and this line of reasoning cannot proceed. However, this is the \emph{only} obstacle, as the homotopy groups
\be
\pi_2(\sdiffthree/H),\quad \pi_1(H)
\ee
are in fact \emph{identical} to the corresponding groups for $G=SO(4)$! We prove this assertion in appendix \ref{appendix_instantons}.

Let us conclude this section with a comparison to the ${\rm SDiff}(S^2)$ super Yang-Mills theory of \cite{Bandos:2008jv}. This can be seen as an $n\to \infty$ limit of $SU(n)$ super Yang-Mills \cite{Hoppe:1988gk}: As observed in \cite{Hitchin},
\be
{\rm SDiff}(S^2)=\mathbf{1}\oplus \mathbf{2} \oplus \mathbf{3} \oplus\dots
\ee
under $SU(2)$ while the expansion for $SU(n)$ (under the adjoint action of $SU(2)$) terminates at spin $(n-1)$. ${\rm SDiff}(S^2)$ super Yang-Mills is an ordinary Yang-Mills theory, albeit with an infinite gauge group and written in different notation. Its vacuum moduli space is given by seven scalars $\phi^{\cal I}, {\cal I}=1,\dots 7$ and the (dual of) the ${\rm SDiff}(S^2)$ gauge field, $A$, all functions on the 2-sphere. The vacuum is supersymmetric if all of these commute
\be
\{\phi^{\cal I},\phi^{\cal J}\} = \{\phi^{\cal I},A\}=0\,,
\ee
where the bracket is the Poisson bracket of functions on $S^2$. The 8 functions $(\phi^{\cal I},A)$ must therefore only depend on a single coordinate, tracing out a curve in $\mathbb{R}^7\times S^1$. The infrared limit (where $S^1$ blows up to $\mathbb{R}$) thus fails to match any of the generic sectors (nonvanishing Hopf invariant) of the $\sdiffthree$ BLG vacuum moduli space.

\section{Discussion}
We have demonstrated that $\sdiffthree$ NCS theory is a higher-spin gauge theory on Euclidean de Sitter space in dimension three, whose gauge group is simple, and which may be seen as a unification of all $\mathfrak{su}(n)\oplus \mathfrak{su}(n)$ theories at once. While that might be an interesting fact in itself (e.g. for cosmological applications), one wonders whether a generalisation to (Euclidean) \emph{anti} de Sitter exists. The obvious candidate is the NCS theory associated to \emph{hyperbolic 3-space} $\mathbb{H}^3\cong SO(3,1)/SO(3)$. Consider an arbitrary 3-manifold $M^3$. Our arguments in sections (\ref{sect_truncation}) and (\ref{sect_higherspin}) only depend on properties of the $\rm SDiff(M^3)$ structure constants $E_{M^3}^{\alpha \beta \gamma}$, and go through as long as
\be
[ISO(M^3),M]\subseteq M
\ee
where $M$ are the complementary generators to those of $ISO(M^3)$ in (the Lie algebra of) $\rm SDiff(M^3)$. This formula is true if $E_{M^3}^{\alpha \beta \gamma}$ is totally antisymmetric; otherwise, the bracket $[ISO(M^3),ISO(M^3)]$ would fail to close on $ISO(M^3)$. The antisymmetry of $E_{S^3}^{\alpha \beta \gamma}$ is most easily seen from its explicit expression as an integral over $S^3$ in appendix \ref{s3harmonics}. It is not a priori clear whether such formulae make sense as written for noncompact 3-manifolds like $\mathbb{H}^3$, but we expect that the issue will be clarified once the harmonic analysis of (divergence-free) vector fields on these spaces is worked out. $\mathbb{H}^3$ in particular might be tractable since $ISO(\mathbb{H}^3)=SO(3,1)$, which is the most familiar of the Lorentz groups.

It is reasonable to expect that this ${\rm SDiff}(\mathbb{H}^3)$ NCS theory describes a ``pyramid of spins'' (\ref{pyramid_of_spins}) around its $AdS_3$ vacuum. A---perhaps disturbing---feature of this spectrum is the infinite proliferation of spin 2 excitations. One suspects that these should not be interpreted as different kinds of gravitons. In 3D the difference between CS gauge theory and CS gravity is that in the latter case the dreibein (a particular combination of CS gauge fields) is assumed to always be invertible, and in particular the vacuum state is not the one where the gauge fields vanish. In section \ref{sect_higherspin} we implicitly assumed that the $SO(4)$-irreducible spin 2 modes (at the top of the pyramid) indeed give rise to an invertible dreibein, but no such requirement was made of the other spin 2 modes, and indeed the argument of Campoleoni et al. in \cite{Campoleoni:2010zq} which establishes the equivalence of the frame-like formalism to the Fronsdal formalism for free 3D 
higher spins only relies on the invertibility of the background dreibein. One expects that this issue will be clarified if the dual CFT is found; presumably that theory would be invariant under the ${\rm SDiff}(\mathbb{H}^3)$ analogue of a $\cal W$-algebra.

One could go on listing various potential applications of ${\rm SDiff}(\mathbb{H}^3)$ NCS gravity. The most striking difference with the $\sdiffthree$ model we have introduced in this paper is that the ${\rm SDiff}(\mathbb{H}^3)$ theory must---if its details work out as expected---admit BTZ black hole solutions. It would be interesting to see whether the ${\rm SDiff}(\mathbb{H}^3)$ gauge symmetry can resolve spacetime singularities.

Throughout this paper we have treated $\sdiffthree$ gauge theory as a 3D gauge theory and not as some sort of rewriting of a 6D gauge theory. It is natural to expect that the $\sdiffthree$ BLG action can be rewritten as the M5 brane action, and indeed this latter point of view has been taken in a number of works. To our knowledge the full $\sdiffthree$ BLG model has not yet been rewritten in six-dimensional language; \cite{Ho:2008nn} did so for the quadratic theory while \cite{Bandos:2008fr} achieved this for the interacting theory in a ``Carrollian'' limit. If this connection is true, our results could be relevant to the physics of M5 branes; one wonders about the interpretation of the quantization of the NCS level in this context.

Let us conclude with some further discussion of the $G=\sdiffthree$ BLG model. While our calculation of the supersymmetric vacuum moduli space produced a result (\ref{moduli_space}) which has a clear interpretation in terms of M2 brane physics, it is somewhat disconcerting that there is no dependence on the NCS level $k$, and in particular that $k\to \infty$ appears not to be a compactification (which is to say, Yang-Mills) limit. For the $G=SO(4)$ model one can easily obtain Yang-Mills theory through the ``novel Higgs mechanism'' of Mukhi and Papageorgakis \cite{Mukhi:2008ux}. This works for $G=SO(4)\cong SU(2)\times SU(2)/\mathbb{Z}_2$ because one can give a VEV to one of the scalars that preserves the diagonal $SO(3)$ subgroup; this is not a generic point of moduli space. There is an obstruction to doing this in $\sdiffthree$ BLG because there is no obvious choice of scalar VEV that gives some distinguished subgroup, and furthermore there is no obvious subgroup to choose (since $\sdiffthree$ is 
simple and not a product). Furthermore the $\sdiffthree$ BLG moduli space $\cal M$ does not appear to contain the moduli space of the $G=SO(4)$ model except for $k=1$ (where $G=SO(4)$ describes two branes on $\mathbb{R}^8$). One still wonders therefore where this theory fits in the greater M theory picture. Right now it is not clear whether e.g. one should expect an $AdS_4$ bulk dual or an $AdS_7$ one; the latter possibility is correct if the model turns out to describe an M5 brane.\footnote{We thank Juan Maldacena for taking the time to point out the $AdS_7$ possibility in between TASI lectures.} We hope our ``global'' results of appendix \ref{appendix_instantons} will be useful in sorting this out in the future.

\section*{Acknowledgements}
I would like to thank the following for providing helpful correspondence, advice, feedback and/or a shoulder to cry on while this work was in preparation: Goro Ishiki, Tyler Kelly, David Tong, Nick Dorey, Carl Turner, James Gundry, Andrew Singleton, Alec Barns-Graham, Chris Blair, Rahul Jha, Ciaran Hughes, Ariana Stylianidi Christodoulou, Juan Maldacena, Will Jay, Emanuel Malek, Panos Betzios, Michele Zoccali and especially Christoph Wockel. Special thanks goes to Allen Hatcher for very patiently explaining his work and for providing general topological guidance, and to my Ph.D. supervisor Paul Townsend for initial collaboration and continuous good suggestions and support throughout, as well as reading a number of versions of this manuscript. Last but not least, I would like to thank the organisers of TASI 2015 as well as Will and Annie Jay for hospitality while part of this work was completed.

I acknowledge support from the UK Science and Technology Facilities Council (grant ST/L000385/1), from Clare Hall College, Cambridge, and from the Cambridge Trust.

\appendix
\section{$S^3$ harmonics}
\label{s3harmonics}
In this paper we make heavy use of the harmonic expansion for divergence-free vector fields on $S^3$, following the notation and conventions of \cite{Aharony:2005bq}. The harmonics will be defined in terms of the round $S^3$ metric of unit radius, and we will make no distinction between vector and 1-form harmonics as one can use the round $S^3$ metric to go from one to the other. We have implicitly assumed here we are using the volume form derived from the round metric, as we have been doing in the main text. This is not restrictive; \emph{Moser's theorem} \cite{Moser:1965} states that any two volume forms (which integrate to the same total volume) may be mapped into each other by a diffeomorphism, so the different $\sdiffthree$ groups in ${\rm Diff}(S^3)$ are all conjugate to each other.

A divergence-free vector harmonic $\cV_\alpha \equiv \cV_{j_\alpha M_\alpha \epsilon_\alpha; i}$ is specified by three indices $(j_\alpha M_\alpha \epsilon_\alpha)$ which we package into a single greek index $\alpha$. The $i$ index is an $S^3$ index and will usually be dropped for brevity. $j_\alpha\in \mathbb{Z}, j_\alpha \geq 1$ and $\epsilon_\alpha=\pm 1$ specify a representation of the isometry group $SO(4)\cong (SU(2) \times SU(2) )/\mathbb{Z}_2$, and $M_\alpha=(m_\alpha,\tilde{m}_\alpha)$ is a pair of $SU(2)$ indices denoting a particular state in that representation. The pair of (half)integer $SU(2)$ spins $(Q_\alpha, \tilde Q_\alpha)$ specifying the representation are
\be
 (Q_\alpha,\tilde Q_\alpha)\equiv \left(\frac{j_\alpha + \epsilon_\alpha}{2} , \frac{j_\alpha - \epsilon_\alpha}{2}\right)
\ee
and $M_\alpha=(m_\alpha, \tilde m_\alpha)$ takes values
\be
 \qquad m_\alpha= -Q_\alpha, -Q_\alpha + 1, \dots Q_\alpha, \qquad \tilde m _\alpha= - \tilde Q_\alpha, -\tilde Q_\alpha +1, \dots \tilde Q_\alpha\,.
 \ee
The vector harmonics are generally complex. Their conjugates are defined as
\be
\label{vectorconjugate}(\cV_{j_\alpha M_\alpha\epsilon_\alpha})^*=(-1)^{m_\alpha+\tilde{m}_\alpha+1}\cV_{j_\alpha (-M_\alpha)\epsilon_\alpha} \\
\ee
We have the orthnormality conditions (Note: vol $S^3$=$2 \pi^2$):
\be
\oint d^3 \sigma e \:(\cV_{j_\alpha M_\alpha\epsilon_\alpha;i})^* \cV_{j_\beta M_\beta\epsilon_\beta;i}
=\delta_{\epsilon_\alpha\epsilon_\beta}\delta_{j_\alpha j_\beta}\delta_{M_\alpha M_\beta}
\ee
or
\be
\oint d^3 \sigma e \: \cV_{\bar{\alpha}} \cV_\beta = \delta_{\alpha \beta}
\ee
where we introduced the barred index notation $\cV_{\bar{\alpha}}\equiv (\cV_\alpha)^*$. We also have the relations
\bea
\nabla_i \cV_{j_\alpha M_\alpha \epsilon_\alpha; i}&=& 0 \\
\label{curleigen}\epsilon_{i j k} \nabla_j \cV_{j_\alpha M_\alpha \epsilon_\alpha; k}&=& - \epsilon_\alpha (j_\alpha+1) \cV_{j_\alpha M_\alpha \epsilon_\alpha; i}
\eea
which is to say, the vector harmonics are divergence-free and are also eigenfunctions of the curl operator associated to the round $S^3$ metric.

We will take a moment to define the coefficients $E^{\alpha \beta \gamma}$ which will turn out to be the structure constants of $\rm SDiff$:
\begin{align}
\label{cubicvectordef} E^{\alpha \beta \gamma} &\equiv \oint  d^3 \sigma e \: \epsilon_{i j k} \cV_{i\alpha} \cV_{j\beta} \cV_{k \gamma} \qquad \text{or} \\ 
E^{j_\alpha M_\alpha\epsilon_\alpha\;j_\beta M_\beta\epsilon_\beta\;j_\gamma M_\gamma \epsilon_\gamma}
&\equiv\oint  d^3 \sigma e \: \epsilon_{ijk}\: 
\cV_{j_\alpha M_\alpha \epsilon_\alpha;i}\cV_{j_\beta M_\beta\epsilon_\beta;j} \cV_{j_\gamma M_\gamma \epsilon_\gamma;k}\,.
\end{align}
$E^{\alpha \beta \gamma}$ are manifestly antisymmetric. They may be calculated from \cite{Aharony:2005bq}:
\begin{align}
\label{cubicvectorcoefficient}
E^{\alpha \beta
\gamma} &=R_{4 \epsilon_\alpha \epsilon_\beta \epsilon_\gamma}(j_\alpha,j_\beta,j_\gamma)\begin{pmatrix} Q_\alpha & Q_\beta &Q_\gamma \\ m_\alpha & m_\beta & m_\gamma \end{pmatrix}\begin{pmatrix} \tilde{Q}_\alpha & \tilde{Q}_\beta & \tilde{Q}_\gamma \\ \tilde{m}_\alpha & \tilde{m}_\beta & \tilde{m}_\gamma \end{pmatrix}
\end{align}
That is two Wigner $3j$ symbols times a reduced matrix element $R_4$. The arguments of nonvanishing symbols satisfy triangle inequalities in the 3 elements of the first row:
\be
\label{triangle}
|Q_\alpha - Q_\beta| \leq Q_\gamma \leq Q_\alpha+Q_\beta, \qquad \text{etc.}
\ee
The numerical coefficient $R_4$ has an explicit expression:
\begin{align}
R_{4 \epsilon_x \epsilon_y\epsilon_z}(x,y,z)=\frac{(-1)^{\sigma'+1}}{\pi}{\rm sgn}(\epsilon_x+\epsilon_y+\epsilon_z) \sqrt{(\sigma'+1)(\sigma'-x)(\sigma'-y) (\sigma'-z) \over
4(x+1)(y+1)(z+1)}\nonumber \\\sqrt{(\epsilon_x(x+1) +
\epsilon_y(y+1) + \epsilon_z(z+1) +2 ) (\epsilon_x(x+1) + \epsilon_y(y+1) + \epsilon_z (z+1)-2)}
\end{align}
The right-hand side of the equation above is defined to be
non-zero only if the triangle inequality 
\be
\label{triangle2}
|x-z| \le y \le x+z
\ee
holds, and if $\sigma' \equiv
(x+y+z+1)/2$ is an integer.

\subsection{$\sdiffthree$ structure constants}
Consider two volume-preserving $S^3$ vector fields $t$ and $s$. The Lie bracket $[t,s]$ can be written
\be
\label{sdiffbracket}
e[t,s]^i= - \varepsilon^{i j k} \partial_j \left(\epsilon_{k l m} t^l s^m\right)\equiv  - \varepsilon^{i j k}\partial_j\left( t \times s\right)_k \,.
\ee
This expression is clearly divergence-free. Let us now define $[t, s]_{j M \epsilon}$ and $(t \times s)_{j M \epsilon}$ by
\be
[t, s]=\sum_{j M \epsilon} [t, s]_{j M \epsilon} \cV_{j M \epsilon}, \qquad t \times s =\sum_{j M \epsilon} (t \times s)_{j M \epsilon} \cV_{j M \epsilon}\,.
\ee
Since the first expression is minus the curl of the second, we can use (\ref{curleigen}) to relate the coefficients:
\be
[t, s]_{j M \epsilon} =  \epsilon (j+1)  (t \times s)_{j M \epsilon}\,.
\ee
What remains is to calculate $ (t \times s)_{j M \epsilon}$. The orthonormality relation gives
\be
 (t \times s)_{j M \epsilon}= \oint d^3 \sigma e (t \times s)^i (\cV_{j M \epsilon;i})^*\,.
\ee
To proceed we select $t= \cV_{j_t M_t \epsilon_t}$, $s= \cV_{j_s M_s \epsilon_s}$ bending our convention on vector harmonic indices slightly to avoid clutter. Then using ({\ref{vectorconjugate}) and (\ref{cubicvectordef}):
\be
\left(\cV_{j_t M_t \epsilon_t}\times \cV_{j_s M_s \epsilon_s}\right)_{j M \epsilon}= (-1)^{m+\tilde{m}+1} E^{j (-M)\epsilon \; j_t M_t\epsilon_t \; j_sM_s\epsilon_s}
\ee
and finally
\be
\label{harmbracket}
[\cV_{j_t M_t \epsilon_t}, \cV_{j_s M_s \epsilon_s}]_{j M \epsilon}=  \epsilon (j+1) (-1)^{m+\tilde{m}+1} E^{j (-M)\epsilon \; j_t M_t\epsilon_t \; j_sM_s\epsilon_s}\, ,
\ee
or in the more compact notation using barred indices
\be
[\cV_t, \cV_s]_\alpha = \epsilon_\alpha (j_\alpha +1) E^{\bar{\alpha} t s}\,.
\ee

Therefore the coefficients $E^{\alpha\beta\gamma}$ defined previously are the structure constants of the Lie algebra ({\it not} the 3-algebra) of ${\rm SDiff}(S^3)$. $E^{\alpha\beta\gamma}$ with the minimal value of $j_\alpha=1$ is especially interesting. When $j_\alpha =1$, one of $Q_\alpha, \tilde Q_\alpha$ vanishes. Using the two triangle inequalities (\ref{triangle}) and (\ref{triangle2}) it is then trivial to show that {\it $E^{\alpha \beta \gamma}$ with $j_\alpha=1$ is only nonzero when $j_\beta=j_\gamma$ and $\epsilon_\beta=\epsilon_\gamma$.} To highlight these facts we will write $E^{\alpha\beta\gamma}$ with $j_\alpha=1$ as
\be
\label{f}
f^\alpha_{\beta \gamma}\equiv f^{M_\alpha\epsilon_\alpha}_{j_\beta M_\beta \epsilon_\beta M_\gamma} \equiv E^{[(j_\alpha=1) M_\alpha \epsilon_\alpha] \: [j_\beta M_\beta \epsilon_\beta] \: [(j_\gamma=j_\beta) M_\gamma (\epsilon_\gamma=\epsilon_\beta)]}\,.
\ee
$f^\alpha_{\beta \gamma}$ is antisymmetric in its lower two indices (which is to say in $M_\beta$ and $M_\gamma$).

\section{The topological classification of $\sdiffthree$ instantons}
\label{appendix_instantons}
In the calculation of the BLG vacuum moduli space for whatever gauge group $G$ a crucial role is played by the periodicity of the scalar that is dual to the unbroken gauge field; integrating out the scalar along with the broken gauge field has the effect of {\it orbifolding} the directions transverse to the M2 branes. The periodicity of the scalar is determined by the quantization of magnetic charge carried by $G$-instantons, which is itself determined by the second homotopy group
\be
\pi_2(G/H)
\ee
where $G$ is the full gauge group and $H$ is the unbroken gauge group, i.e. the stabiliser of any point of moduli space ${\cal M}(\sdiffthree {\rm BLG})$ as defined in the main text. This much has been detailed in section \ref{sect_vacuummoduli} of this paper and the original references \cite{Lambert:2008et,Distler:2008mk,Lambert:2010ji}. Here we will calculate $\pi_2(G/H)$ in the case $G={\rm SDiff}(S^3)$. Our main tool will be the long exact homotopy sequence of the fibration (fibre bundle) $G \to G/H$, so we will begin by calculating the homotopy type of $G={\rm SDiff}(S^3)$ and $H$.

We will get a handle on the topology of ${\rm SDiff}(S^3)$ using the famous result of Hatcher \cite{Hatcher:1983} which states
\begin{theorem}
The group ${\rm Diff}(S^3)$ of diffeomorphisms of the 3-sphere $S^3$ strongly deformation retracts onto its $O(4)$ subgroup.
\end{theorem}
In particular, the two groups have the same homotopy type. This does not immediately descend to the group of {\it volume-preserving} diffeomorphisms ${\rm SDiff}(S^3)$ that we are interested in\footnote{I would like to thank Allen Hatcher for pointing this out to me.}. However, the relationship between them is known. The following result was originally proven by Ebin and Marsden in the context of geometric hydrodynamics: 
\begin{theorem}
{\rm(5.1) of \cite{Ebin:1970}.} For any compact orientable manifold $M$ without boundary,
\begin{align*}
{\rm Diff}(M) \qquad \text{is diffeomorphic to} \qquad {\rm SDiff}(M)\times {\cal V}
\end{align*}
where ${\cal V}$ is the space of volume forms of $M$ of (any) {\it fixed} total volume, i.e.
\begin{align*}
\forall \mu 1,\mu 2 \in {\cal V},\qquad \int_M \mu 1=\int_M \mu 2\neq 0\,.
\end{align*}
\end{theorem}
This theorem is perhaps most easily understood as the statement that the principal bundle ${\rm Diff}(M)/{\rm SDiff}(M)$ is trivial, and may be identified with the space $\cal V$ (Moser's theorem \cite{Moser:1965}).

${\cal V}$ is an affine space, whence contractible, yielding a (strong) deformation retraction of ${\rm Diff}(S^3)$ to its ${\rm SDiff}(S^3)$ subgroup. Combined with Hatcher's result, we see the following
\begin{theorem}
$SO(4)$ and $\sdiffthree$ are homotopy equivalent; they are both deformation retracts of ${\rm Diff}(S^3)$.
\end{theorem}
Let us reiterate here that we always consider the identity component of any diffeomorphism groups in this paper unless otherwise stated. This theorem implies in particular
\bea
\pi_1({\rm SDiff}(S^3))&=& \mathbb{Z}_2\,, \\
\pi_2({\rm SDiff}(S^3))&=& 0\,.
\eea

Let us give a less terse definition of the unbroken gauge group $H$ here. If $h: S^3 \to S^2$ is the Hopf map that makes $S^3$ into the total space of an $U(1)$-bundle ${\cal P}$ over $S^2$, we have
\begin{align}
g\in H \subseteq {\rm SDiff}(S^3) \iff h\circ g = h\,.
\end{align}
It is not hard to see that this is the set of $g\in \sdiffthree$ that act trivially on $\vec{\phi}:S^3 \to \mathbb{R}^8$ which make the 3-bracket vanish, i.e. points on ${\cal M}(\sdiffthree {\rm BLG})$. The key point is that $g\in H$ thus defined fixes the $S^2$ base and is volume-preserving on the $S^3$ total space, hence it is volume-preserving on each $S^1$ fibre. But this means it acts as a $U(1)$ transformation on the fibres! $H$ is thus equivalently described as the group of {\it gauge transformations} ${\rm Gau}({\cal P})$ of the U(1)-bundle $\cal P$, also known as vertical automorphisms in the mathematical literature. There exists a result by Wockel which yields information about the homotopy type of ${\rm Gau}({\cal P})$, valid under rather general conditions:
\begin{lemma}
{\rm (II.8.) of \cite{Wockel:2006}.} If ${\cal P} = (K, h: P\to S^m)$ is a principal $K$-bundle over $S^m$ and $K$ is locally contractible, there is a long exact homotopy sequence
\be
\label{wockelsequence}
\dots \to \pi_{n+1}(K) \to \pi_{n+m}(K) \to\pi_n({\rm Gau}({\cal P}))\to \pi_n(K) \to \pi_{n+m-1}(K) \to \dots
\ee
\end{lemma}
This homotopy sequence arises from the continuous fibre bundle ${\rm Gau}({\cal P})\xrightarrow{{\rm ev}_{x_0}}K$, see \cite{Wockel:2006} for details. Here we have $m=2$, $K=U(1)\cong S^1$, and ${\rm Gau}({\cal P})=H$ and the long exact homotopy sequence becomes
\be
\dots \to \pi_{n+2}(S^1) \to \pi_n(H) \to \pi_n(S^1)\to \dots
\ee
implying
\bea
\pi_0(H)&=&0 \\
\pi_1(H) &=& \mathbb{Z} \\
\pi_n(H) &=& 0 \qquad\forall n\geq 2 \,.
\eea
$H$ is therefore a (weak) homotopy $U(1)$.

We are now in a position to calculate the homotopy groups of interest. We have the long exact homotopy sequence associated to the principal fibre bundle ($G={\rm SDiff}(S^3)$)
\be
H\xrightarrow{\iota} G \to G/H\,,
\ee
viz.
\be
\dots\to\pi_2(G) \to \pi_2(G/H) \to\pi_1(H) \to \pi_1(G) \to \dots
\ee
Using the previous results this becomes
\be
\dots \to 0\to \pi_2(G/H) \xrightarrow{\rho} \mathbb{Z} \xrightarrow{\sigma}\mathbb{Z}_2\to \dots
\ee
At this point we remind the reader that the arrows in an exact sequence are group homomorphisms such that the kernel of each arrow is the image of the preceding one. Therefore
\begin{align}
\pi_2(G/H)&= \ker(\sigma)\cong\mathbb{Z}\,.
\end{align}

There is an ambiguity in that $\pi_2(G/H)$ could be either $\mathbb{Z}$ or $2\mathbb{Z}$, corresponding to whether $\sigma$ is the zero map or not. This is easily resolved however: Consider the commutative square
\be
\begin{CD}
H @<< i<U(1)\\
@VV\iota V @VV d V\\
G @<<< SO(4)
\end{CD}
\ee
where $d$ is the ``diagonal'' embedding of $U(1)$ in $SO(4) \cong (SU(2) \times SU(2)) / \mathbb{Z}_2$, $\iota$ is the inclusion of $H$ into $G$ which induces the map $\sigma$ between $\pi_1(H)$ and $\pi_1(G)$, and the horizontal arrows are also inclusions; $i$ in particular is the $U(1)$ action on $S^3$ that makes it into a fibre bundle. $d$ clearly induces the zero map on the fundamental groups\footnote{Let $x\in SU(2)\cong S^3$, then $SO(4)$ acts by left and right multiplication by $SU(2)$ elements $(g_L,g_R)$ and we have $(-g_L,-g_R)\sim (g_L,g_R)$. If $g_L=g_R=e^{i \phi}$ we have $(e^{i\phi},e^{i\phi}) \sim(e^{2i\phi},1)\sim (1,e^{2 i \phi})$. This gives the trivial element of $\pi_1(SO(4))=\mathbb{Z}_2.$}. The inclusion $i:U(1) \to H$ induces an isomorphism between the fundamental groups; this is because the map ${\rm ev}_{x_0}: H \to U(1)$ which induces the sequence (\ref{wockelsequence}) of reference \cite{Wockel:2006} also induces an isomorphism (we proved this below Lemma 1), and because ${\rm ev}_{
x_0}\circ i = {\rm id}_{U(1)}$ since ${\rm ev}_{x_0}$ is evaluation of 
an element of $H$---seen as a space of \
\emph{equivariant} mappings to the gauge group $U(1)$ of the Hopf bundle---at some point $x_0$ on $S^3$. Now the induced square on the fundamental groups $\pi_1$ also commutes and we therefore have
\be
\sigma i_\ast =  0 \implies \sigma =0,
\ee
the zero map.

\end{document}